# Unconventional spin textures emerging from a universal symmetry theory of spin-momentum locking

Yuntian Liu[1], Jiayu Li[1], Pengfei Liu[1] and Qihang Liu[1,2,3,*]

[1]*Shenzhen Institute for Quantum Science and Engineering and Department of Physics, Southern University of Science and Technology, Shenzhen 518055, China*

[2]*Guangdong Provincial Key Laboratory of Computational Science and Material Design, Southern University of Science and Technology, Shenzhen 518055, China*

[3]*Shenzhen Key Laboratory of Advanced Quantum Functional Materials and Devices, Southern University of Science and Technology, Shenzhen 518055, China*

[*]Email: liuqh@sustech.edu.cn

**Abstract**

Spin textures, i.e., the distribution of spin polarization vectors in reciprocal space, exhibit diverse patterns determined by symmetry constraints, resulting in a variety of spintronic phenomena. Here, we propose a universal theory to comprehensively describe the nature of spin textures by incorporating three symmetry flavors of reciprocal wavevector, atomic orbital and atomic site. Such approach enables us to establish a complete classification of spin textures constrained by the little co-group and predict unprecedentedly reported spin texture types, such as Zeeman-type spin splitting in antiferromagnets and quadratic spin texture. To illustrate the influence of atomic orbitals and sites on spin textures, we predict orbital-dependent spin texture and anisotropic spin-momentum-site locking effects, and corresponding material candidates validated through first-principles calculations. The comprehensive classification and the predicted new spin textures in realistic materials are expected to trigger future spin-based functionalities in electronics.

*Introduction*

Spintronic effects, based on the polarization of electronic spin, hold the potential to develop the next-generation devices with low energy consumption and ultrafast processing speeds[1-4]. While in ferromagnetic materials the itinerant electrons are spin polarized by the magnetic ions, in nonmagnetic systems with sufficiently low symmetry, relativistic spin-orbit coupling (SOC) can also generate an effective magnetic field that couples spin to momentum[5], leading to a distinct type of spin polarization named spin-momentum locking. Specifically, such momentum-dependent spin polarization is the expectation value of spin operators $\vec{\sigma} = (\sigma_x, \sigma_y, \sigma_z)$ in a given Bloch wave function $|u_n(\boldsymbol{k})>$, resulting in a vector texture $\vec{S}_{nk} = \langle u_n(\boldsymbol{k})|\vec{\sigma}|u_n(\boldsymbol{k})\rangle$ throughout the Brillouin zone, known as spin texture. The diversity of spin texture provides a prospect for the rich spintronic applications, such as spin field-effect transistors[6,7], spin-galvanic effects[8], spin Hall effect[3,9,10], spin-orbit torque[11-13], etc.

Designing spin textures by symmetry is an effective approach for material selection. Historically, the typical Dresselhaus[14] and Rashba[15] spin textures were believed to arise from the bulk and structural inversion asymmetry of the material, respectively[5]. However, such loose symmetry classification has been considered incomplete with the subsequential discovery of new spin textures, including Rashba spin splitting in bulk materials[16], cubic Rashba spin splitting[17], Zeeman-type spin splitting in nonmagnetic materials[18,19], and persistent spin textures[20-22]. On the other hand, spatially localized spin polarization could also arise in materials with $PT$ symmetry (the combination of space-inversion $P$ and time-reversal $T$) where spin degeneracy exists throughout the whole Brillouin zone[23-28], indicating that the real-space degree of freedom also plays a role in novel spin polarization. These studies show the variety of spin textures yet imply a limited understanding of the ultimate origin of spin textures. Therefore, a symmetry-based universal theory that categorizes all possible spin textures, which would also facilitate the discovery of new spintronic phenomena and materials, is urgently needed.

In this work, we utilize state-of-the-art symmetry theory to comprehensively investigate the symmetry-dictated spin textures by considering three indispensable flavors: little co-group of the wavevector *k* (i.e., the symmetry group that remains *k* unchanged or differs by an integer multiple of reciprocal lattice vectors), atomic site and atomic orbital. First, we provide a complete classification of spin textures under the magnetic point group according to the lowest order of the symmetry-allowed spin-splitting *k*-polynomials of the effective Hamiltonian. Based on such classification, we predict several unconventional spin textures, including quadratic spin texture in nonmagnetic materials and Zeeman-type spin splitting in antiferromagnetic (AFM) materials. Furthermore, we considered the effects of additional atomic orbital and site degrees of freedom on spin textures, leading to orbital-dependent spin texture and spin-momentum-site locking, respectively. Our work provides a complete understanding of the interplay between spin textures and symmetries, paving an avenue offering future possibilities for spintronic effects.

*Methodology*

To comprehensively capture the possible scenarios of spin textures, we utilize the group decomposition method involving three related ingredients, i.e., specific *k*-points, atomic orbitals and Wyckoff positions, as illustrated in Fig. 1. To begin with, we adopt the magnetic space group $G$, atomic orbitals $i$, the occupied Wyckoff positions $q$ of a given material and the considered wavevector $k$, as input for the whole process. Under the mean-field approximation, such decomposition methodology is sufficient to cover all possible types of spin textures.

Next, we construct the corresponding band representations $\rho^k$ of the little co-group $G^k$ using Zak's method[29]. Specifically, the site symmetry representation $\rho^q$ of the atomic orbitals in real space is induced to the global magnetic space group representation $\rho^q \uparrow G$, and then subduced to the little co-group of the selected *k*-point in reciprocal space $\rho^k = (\rho^q \uparrow G) \downarrow G^k$. For the spinful system, the complete representation requires the spinor representation $U$ on the direct product, written as $\rho^k \otimes U^k$.

Finally, we derive the $k \cdot p$ Hamiltonian near a specific wavevector $k$ with little co-group $G^k$. The Hamiltonian is then solved to determine wavefunctions, band structures, and spin textures with appropriate undetermined coefficients. Therefore, the wavefunctions that contain the information of atomic orbitals and Wyckoff positions can be obtained, enabling a complete depiction of the spin textures including the information of both orbital and site degrees of freedom. Under the mean-field approximation, such methodology is sufficient to cover all possible types of spin textures. We provide a detailed derivation of the representation matrices and $k \cdot p$ Hamiltonian in the Supplementary Note I.

*Spin textures classified by k-polynomials*

While it is recently realized that the symmetry of the specific *k*-point plays a crucial role in determining the spin texture[30], a full classification of spin-momentum locking constrained by little co-group, which is favorable to predict new effects beyond nonmagnetic point group symmetry and linear spin splitting, is still lacking. From the perspective of the low-energy effective Hamiltonian, the spin texture is largely determined by the order of the spin-splitting *k*-polynomial (e.g., $k_i^n \sigma_j$ is an $n$-order spin-splitting term with $\sigma_j$ a Pauli matrix for spin), such as 0th-order (Zeeman-type); 1st-order (Dresselhaus-, Rashba-, and Weyl-type) and 3rd-order (cubic Rashba-type). Thus, we consider the $k \cdot p$ Hamiltonian near a *k*-point with all the magnetic point group lacking $PT$ symmetry (the combination of space-inversion P and time-reversal T) as its little co-group and classify the corresponding spin texture according to the lowest order of the *k*-polynomial (typically the dominant term), as shown in Table I.

We next discuss the symmetry conditions for each type of spin texture. Firstly, $T$ breaking and $P$ breaking are essential prerequisites for the emergence of even-order and odd-order spin textures, respectively. Therefore, typical time-reversal invariant momenta in nonmagnetic materials exclusively exhibit 1st-order or 3rd-order spin textures. Next, we provide specific symmetry requirements for the spin textures in each order. Besides $T$ or $P$ breaking, the 0th-order Zeeman-type spin textures require breaking $3_{111}$, $T4_{001}$, $T6_{001}$, $PT4_{001}$, and $PT6_{001}$, with only one $n$ or $\bar{n}$ rotation axis. For the 1st-order spin textures, one needs to break $P6_{001}$, $PT6_{001}$, the combination of $\{P4_{001}, 3_{111}\}$, or $\{PT4_{001}, 3_{111}\}$ (the bracket means that one of $PT4_{001}$ and $3_{111}$ needs to be broken). Additionally, certain conditions are required to prevent lower-order (e.g., 0th-order) spin-splitting terms, such as the presence of $3_{111}$, $T4_{001}$, $T6_{001}$, $PT4_{001}$, or possessing rotation operations $n$ and $\bar{n}$ along multiple directions (named "multi-axis rotation"). Similarly, the 2nd-order spin textures require breaking the combination of $\{P4_{001}, 3_{111}\}$ and $\{T, P6_{001}\}$, while excluding the conditions supporting the 0th- and 1st-order ones. The 3rd-order spin textures correspond to only two cases, i.e., the combination of $\{P4_{001}, 3_{111}\}$ or $\{T, P6_{001}\}$, leading to four possible magnetic point groups. Lastly, the 4th-order spin textures can be supported by only one magnetic point group $m\bar{3}m$. These symmetry conditions result in the second column of Table I, facilitating the prediction of new materials with unconventional spin textures.

*Quadratic Spin textures*

Based on Table I, several new spin textures are emerging. Firstly, the highest order of the dominant spin texture in magnetic group is quartic (i.e., the lowest spin-splitting terms are 4th-order $k$-polynomial $H_{split} = k_x k_y(k_x^2 - k_y^2)\sigma_z + k_y k_z(k_y^2 - k_z^2)\sigma_x + k_x k_z(k_z^2 - k_x^2)\sigma_y$), where only one little co-group ($m\bar{3}m$) is available in certain complex cubic magnets. Another novel spin texture type is quadratic spin texture, which has generally been overlooked in nonmagnetic materials. Table I lists the collection of 15 magnetic point groups hosting quadratic spin textures, and only three of them ($\bar{6}', \bar{6}m2, \bar{6}'m2'$) can appear in nonmagnetic materials. Further symmetry analysis reveals that the quadratic spin texture requires $P6_{001}$ or $PT6_{001}$ with broken time-reversal $T$. In comparison, it is easier to achieve quadratic spin textures in magnetic materials, which only requires space-inversion $P$ to forbid the 1st-order terms and symmetries that prohibiting the 0th-order terms.

Next, we propose realistic candidate materials to demonstrate quadratic spin textures by employing first-principles calculations (Supplementary Note II). We choose BaSbPt as an example, which has a hexagonal lattice with type-II magnetic space group $P\bar{6}m21'$ and lattice constants $a = b = 4.568$ Å, $c = 4.986$ Å[31] with the Ba, Sb, Pt atoms occupying Wyckoff positions 1a, 1f, 1d, respectively (Fig. 2a). The little co-group of the high-symmetry line Γ-$A$ is $\bar{6}'m2'$, which belongs to quadratic spin texture according to our classification. The $k \cdot p$ spin-splitting Hamiltonian is written as $H_{split} = (k_x^2 - k_y^2)\sigma_x - 2k_x k_y \sigma_y$. The spin texture of BaSbPt on the $k_z = 0.25$ plane is shown in Fig. 2b. The in-plane components manifest $\vec{S}(k_x, k_y) = \vec{S}(-k_x, -k_y)$, and thus are totally distinct from those of the conventional nonmagnetic materials (e.g., the linear or cubic Rashba/Dresselhaus spin textures[17].)

*Zeeman spin splitting in antiferromagnets*

From Table I we also find that many familiar spin textures may appear in systems previously thought not to exhibit them. Particularly, the magnetic point groups that exhibit 0th-order terms, except $1$ and $\bar{1}$, could also allow AFM configurations, suggesting that Zeeman-type spin splitting can also occur even at the Γ point of antiferromagnets. Such AFM Zeeman-type spin splitting applies to all of the AFM systems that exhibit the anomalous Hall effect[32,33], as they share the same symmetry requirements.

For material realization, we choose collinear AFM MnTe with Néel vector along the *c*-axis as an example. MnTe has an orthogonal magnetic space group $Pnm'a'$ and lattice constants *a* = 6.753 Å, *b* = 4.171 Å and *c* = 7.237 Å[31] with both of the Mn and Te atoms occupying Wyckoff positions 4c (Fig. 2c). The spin-resolved band structure is shown in Fig. 2d with a significant Zeeman-type spin splitting at the Γ point ($H_{split} \propto \sigma_z$), which was previously considered as a characteristic of ferromagnet. It should be notice that the Zeeman-type splitting in AFM (spin polarization along *x*-axis) could not appear without SOC. This means that such Zeeman-type spin splitting at Γ is totally originated from the SOC effect rather than the altermagnetic spin splitting[34]. In addition to the cases described above, we also present several novel spin textures determined by little co-groups in the Supplementary Note III.

*Spin texture determined by atomic orbitals*

Next, we elucidate the influence of atomic orbitals and sites on spin textures, as well as the emergence of novel spin textures resulting from the coupling among these degrees of freedom. We first consider the effect of atomic orbitals, which has been discussed as "spin-orbital texture" or "spin-orbital-momentum locking"[35,36]. While it is natural to understand from the previous literature that the atomic orbitals serve as the bases of different band representations, leading to the diversity of spin textures, our comprehensive classification indicate that the spin textures are also determined by the dimensions of orbital bases. As we mentioned before, the representation of symmetry operations can be written as $A^k = \rho^k \otimes U^k$ with the basis $(l_1, l_2, ..., l_n) \otimes (\uparrow, \downarrow)$, where $l_i$ is the atomic orbital and $n$ is the dimension of the orbital basis. For single-orbital basis ($n = 1$), the representation of orbital $\rho^k = e^{i\alpha}$ is nothing but a phase factor dependent on the type of the orbital, whose effect on the form of the Hamiltonian will be eliminate by the invariant relation $A^K(g)H^K(\boldsymbol{k})A^K(g)^{-1} = H^K(g\boldsymbol{k})$. Therefore, all the single-orbital bases within a given little group will carry the spin textures with the same form of the $k \cdot p$ Hamiltonian, regardless of the difference in orbital type or band representation.

For multi-orbital bases ($n > 1$), the representation of orbital $\rho^k$ has a matrix form and thus cannot be completely eliminated by the invariant relation. Therefore, multi-orbital bases can support spin textures different from those of single-orbital bases, even if they share the same band representations. Here, we choose the $p_x p_y$ and $s$ orbitals in AuCN as an example, which has a hexagonal lattice with type-II magnetic space group $P6mm1'$ and lattice constants $a = b = 3.662$ Å and $c = 5.113$ Å[31] with all of the Au, C and N atoms occupying Wyckoff positions 1a (Fig. 3a). The calculated band structures show that both the $p_x p_y$ (Fig. 3b) and $s$ orbitals (Fig. 3c) manifest double degeneracy at the $K$ point with the same representation $\bar{K}_6$. However, the spin texture of $p_x p_y$ orbitals is dominated by 2nd-order $k$-polynomial (Fig. 3d), while that of $s$ orbital behaves quite differently, showing a 1st-order Rashba-like feature (Fig. 3e).

We next use our theoretical approach to explain this peculiar phenomenon. Specifically, the wavefunction bases that carry this two-dimensional representation can be written as $(s\uparrow, s\downarrow)$ for $s$ orbital and $(p_+\uparrow, p_-\downarrow)$ for $p$ orbital, where $|p_\pm> =$ $|p_x> \pm i|p_y>$, ↑ and ↓ mean spin up and down. Consequently, the $k \cdot p$ Hamiltonian for the two orbitals can be written as:

$$H^K_{(s\uparrow,s\downarrow)} = h_1(k_y\sigma_x - k_x\sigma_y) + h_2[(k_x^2 - k_y^2)\sigma_y + 2k_xk_y\sigma_x] \quad (1)$$

$$H^K_{(p_+\uparrow,p_-\downarrow)} = h_1(k_y\upsilon_x + k_x\upsilon_y) + h_2[(k_x^2 - k_y^2)\upsilon_y - 2k_xk_y\upsilon_x] \quad (2)$$

where $\sigma$ and $\upsilon$ represent the Pauli matrices with the corresponding basis; $h_i$ are the undetermined coefficients. Both Hamiltonians are expanded up to 2nd-order $k$-polynomial, while the constant and kinetic terms are excluded because they have no impact on the spin texture. Eq. (1) reveals that the $s$-orbital-dominated Hamiltonian primarily manifest the 1st-order Rashba term and accompanied by a 2nd-order warping term, which is consistent with our DFT calculations (Fig. 3e). In comparison, the $p$-orbital-dominated Hamiltonian described by Eq. (2) do not exhibit spin polarization, because the basis $(p_+\uparrow, p_-\downarrow)$ does not correspond to the real spin. Therefore, it is necessary to consider an additional set of $(p_+\downarrow, p_-\uparrow)$ orbitals represented by $\bar{K}_4 \oplus \bar{K}_5$, which do not hybridize with $(p_+\uparrow, p_-\downarrow)$ at the $K$ point. However, away from the $K$ point hybridization between the two sets of $p$ orbitals is permitted, resulting in the emergence of a quadratic spin texture (Fig. 3d) with additional hybridization terms $(k_x\tau_x \otimes \sigma_0 - k_y\tau_y \otimes \sigma_0)$, where $\tau$ and $\sigma$ represent the Pauli matrices with the basis of $(p_+, p_-)$ and $(\uparrow, \downarrow)$, respectively.

It can be seen that the variation in spin textures depend on the type of orbital basis and does not involve the distribution of orbital components in reciprocal space. Therefore, different from the physics spin-orbital texture[35,36], such "orbital-dependent spin texture" offers an additional knob for designing spin textures. In Supplementary Note IV we provide more material examples arising such orbital-dependent spin texture with even identical representation.

*Spin-momentum-site locking effect*

We next focus on the influence of atomic sites on global and site-resolved spin textures, i.e., spin-momentum-site locking. The correlation between atomic sites and spin polarization was initially found in nonmagnetic materials with two space-inversion breaking sectors connected by global space inversion $P$, referred to as "hidden spin polarization"[23-28]. Due to the diversity of the atomic sites, the influence of site freedom on spin texture is not limited to hidden spin polarization, but also encompasses various coupling forms according to our classification procedure.

Here, we predict an anisotropic spin-momentum-site locking effect originated from the coupling between low site symmetry in real space and specific wavevector in momentum space. We demonstrate this phenomenon in GaCuBr$_4$, which has a tetragonal lattice with type-II magnetic space group $P\bar{4}2c1'$ and lattice constants $a = b = 5.785$ Å and $c = 10.629$ Å[31] with the Ga, Cu, Br atoms occupying Wyckoff positions 2b, 2e, 8n, respectively (Fig. 4a). Importantly, Ga atoms have a site symmetry group of $2221'$, which exhibits a strong anisotropy compared with the global symmetry. We select the band mainly consisting of Ga atoms (Fig. 4b) and illustrate its spin texture in the $k_z = 0.5$ plane (Fig. 4c), showing Dresselhaus-type spin texture near all four time-reversal invariant momenta. The spin texture is projected onto two GaBr$_4$ tetrahedron sites $\vec{S}_{A,nk} = \langle u_n(\bm{k})|P_A\vec{\sigma}P_A|u_n(\bm{k})\rangle$ as shown in Fig. 4d-e, where $P_A$ represents the projection operator of the site A. Compared the projected spin textures of two Ga sites with the global spin texture, we find interestingly strong anisotropic distribution that Ga site 1 exhibits a concentrated spin polarization near the $k_y = 0$ line (Fig. 4d), while the other site shows a concentration near the $k_x = 0$ line (Fig. 4e) separated by the $k_x = \pm k_y$ line. These results clearly reveal an anisotropic spin-momentum-site locking phenomenon.

The coupling pattern between spin, momentum and site presents novel opportunities for the manipulation of spin polarization distribution in real space. For example, the degeneracy caused by two symmetry-connected sites distributed along the *z*-direction (such as two Ga sites in GaCuBr$_4$) can be lifted by applying a perpendicular electric field, leading to an anisotropic spin texture dominated by one site. Consequently, the induced spin-polarized current can be confined to a specific direction.

*Discussion*

In general, our work establishes the connections between spin textures and complete symmetries, and, from this perspective, we discover several novel spin textures. We provide comparisons between the spin textures obtained by our models and DFT calculations in the Supplementary Note V to showcase the validity of the symmetry description in predicting spin textures of realistic materials. In addition, this framework can be easily extended to describe other fiber bundle vector fields in reciprocal space, including orbital texture[37,38], spin-orbit texture[35,39], Berry curvature[40,41], etc. Furthermore, our method can also be readily applied to the recently studied SOC-free systems[34,42-44], which are essentially described by spin crystalline groups[45-49]. This can be achieved simply by replacing the Wyckoff positions and little co-groups with their corresponding spin crystalline group counterparts[45-49].

The predicted new spin textures, including multi-degree of freedom coupling, provides potential applications for spintronics. For example, applying an electric field in the *xy*-plane of the quadratic spin textures can generate an unconventional reciprocal Edelstein effect (i.e., changing the sign of the electric field will not flip the spin) with the spin polarization pinned to the electric field direction, which could be benefit for spin-orbit torque applications. For the orbital-dependent spin textures, polarized light can be utilized to select atomic orbitals with specific angular momenta, thus manipulating the corresponding spin textures. In addition, anisotropic spin-momentum-site locking effect offer the opportunity to modify the strength of current-induced spin polarization by an electric field perpendicularly to the current. Therefore, we expect that the exploration of these novel spin textures and their versatile applications in spintronics will pave the way for the advancements of future electronic devices.

**Acknowledgements**

This work was supported by National Natural Science Foundation of China under Grant No. 12274194, Guangdong Provincial Key Laboratory for Computational Science and Material Design under Grant No. 2019B030301001, Shenzhen Science and Technology Program (Grant No. RCJC20221008092722009), the Science, Technology and Innovation Commission of Shenzhen Municipality (Grant No.

**Table I | Classification of spin texture according to the lowest order of $k$-polynomial of $k \cdot p$ Hamiltonian allowed by the magnetic point groups of $k$-points.** Red fonts mark the magnetic point groups that can only appear in magnetic systems. The magnetic point groups are denoted by international symbols. Time-reversal symmetry and space-inversion symmetry are denoted as $T$ and $P$, respectively. The material examples are also provided.

| Lowest order | Spin texture | Magnetic point groups | Brief symmetry condition | Material examples |
|---|---|---|---|---|
| 0 | Zeeman | $1, \bar{1}, 2, 2', m, m', 2/m, 2'/m',$ $2'2'2, m'm2', m'm'2, m'm'm,$ $4, \bar{4}, 4/m, 42'2', 4m'm',$ $4/mm'm', 3, \bar{3}, 32', 3m', \bar{3}m',$ $6, \bar{6}, 6/m, 62'2', \bar{6}m'2',$ $6/mm'm'$ | Break $T$ | $MoSe_2^{18}$, MnTe |
| 1 | Linear | $11', 21', m1', 222, 2221', mm2,$ $mm21', 41', 4', \bar{4}1', \bar{4}', 422,$ $4221', 4'22', 4mm, 4mm1',$ $4'm'm, \bar{4}2m, \bar{4}2m1', \bar{4}'2'm,$ $\bar{4}'2m', \bar{4}2'm', 31', 32, 321',$ $3m, 3m1', 61', 6', 622, 6221',$ $6'22', 6mm, 6mm1', 6'mm',$ $6m'm', 23, 231', 432, 4321',$ $4'32'$ | Break $P$ | $Bi/Ag^{50}$, $GaAs^{51}$, $BiTeI^{16}$ |
| 2 | Quadratic | $mmm, 4'/m, 4/mmm,$ $4'/mm'm, \bar{3}m, \bar{6}', 6'/m', \bar{6}m2,$ $\bar{6}'m'2, \bar{6}'m2', 6/mmm,$ $6'/m'mm', m\bar{3}, \bar{4}'3m', m\bar{3}m'$ | Break $T$ | BaSbPt |
| 3 | Cubic | $\bar{6}1', \bar{6}m21',$ $\bar{4}3m, \bar{4}3m1'$ | Break $P$ | $Ge_3Pb_5O_{11}^{17}$ |
| 4 | Quartic | $m\bar{3}m$ | Break $T$ | -- |

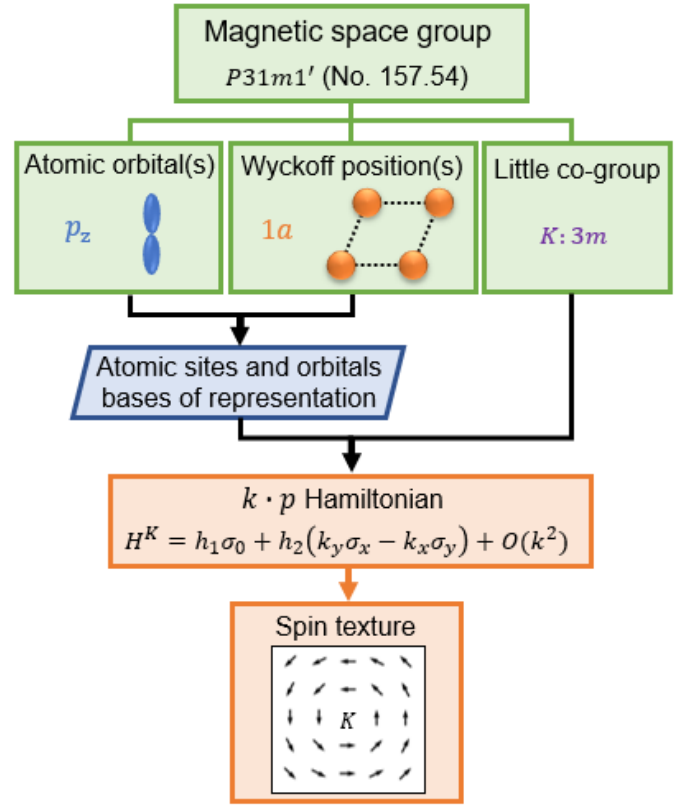

**Fig. 1. | Procedures for comprehensively obtaining spin textures by using group decomposition method involving three ingredients.** As an example, a compound with magnetic space group $P31m1'$ is considered with $p_z$ orbital located at Wyckoff position $1a$ (top green boxes) around the Fermi level. The obtained $k \cdot p$ Hamiltonian near the $K$ point and the resultant Rashba spin texture are displayed in the bottom orange boxes.

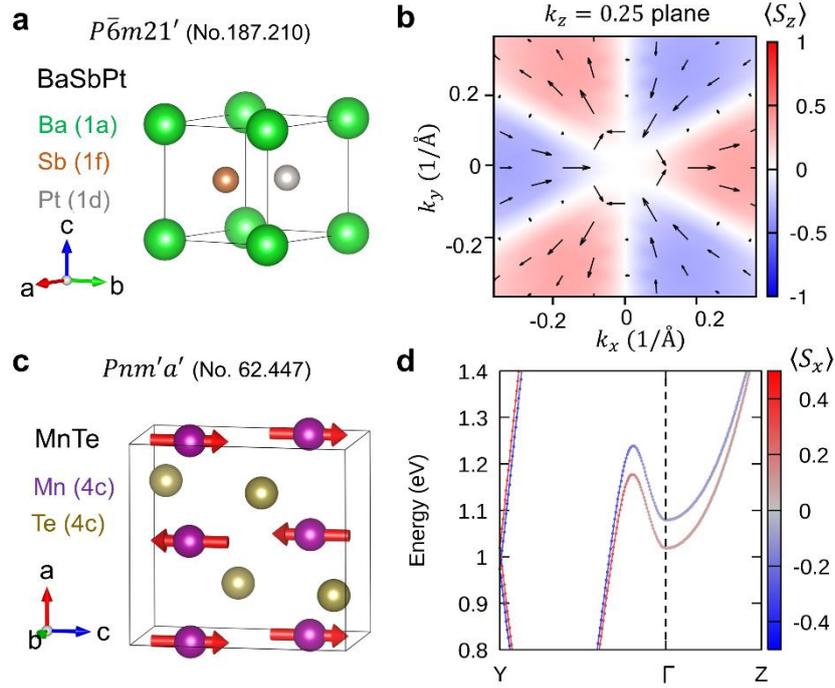

**Fig. 2 | Quadratic spin texture in BaSbPt and Zeeman-type spin splitting in AFM MnTe. a** Crystal structure and symmetry information of BaSbPt. **b** The quadratic spin texture of BaSbPt centered at the Γ-$A$ line on the $k_z = 0.25$ plane. The arrows represent the in-plane spin polarization and the background colors represent the out-of-plane component. **c** Crystal structure and symmetry information of MnTe. **d** Spin-resolved band structure of MnTe, showing Zeeman-type spin splitting at the Γ point.

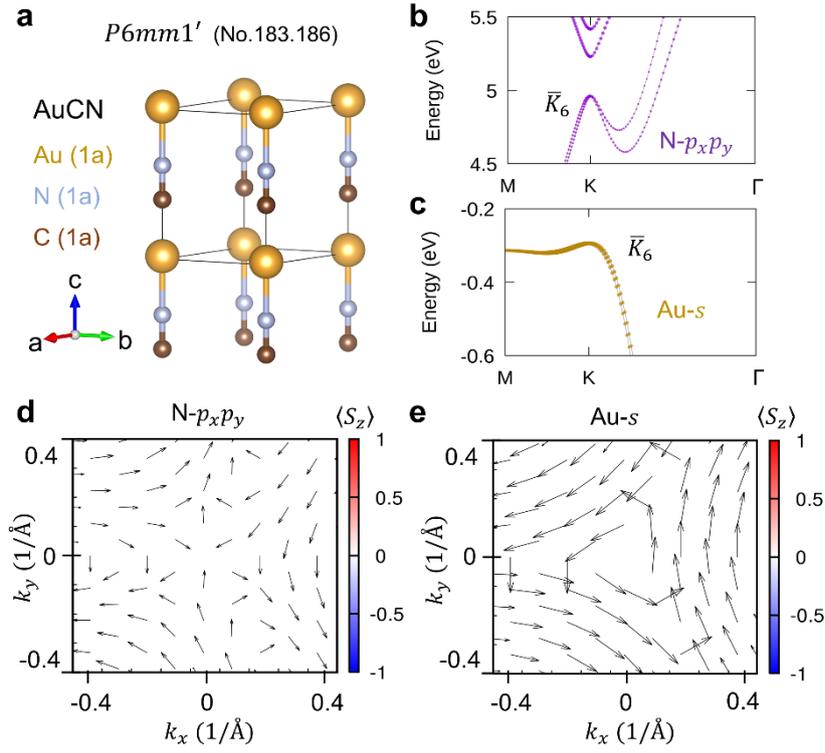

**Fig. 3 | Orbital-dependent spin texture in AuCN. a** Crystal structure and symmetry information of AuCN. **b-c** Band structures at the *K* valley of **b** $p_x p_y$ orbitals and **c** $s$ orbital in AuCN. **d-e** Spin textures of **d** $p_x p_y$ orbitals and **e** $s$ orbital centered at the *K* valley in AuCN. The arrows represent the in-plane spin polarization, and the background colors represent the out-of-plane component.

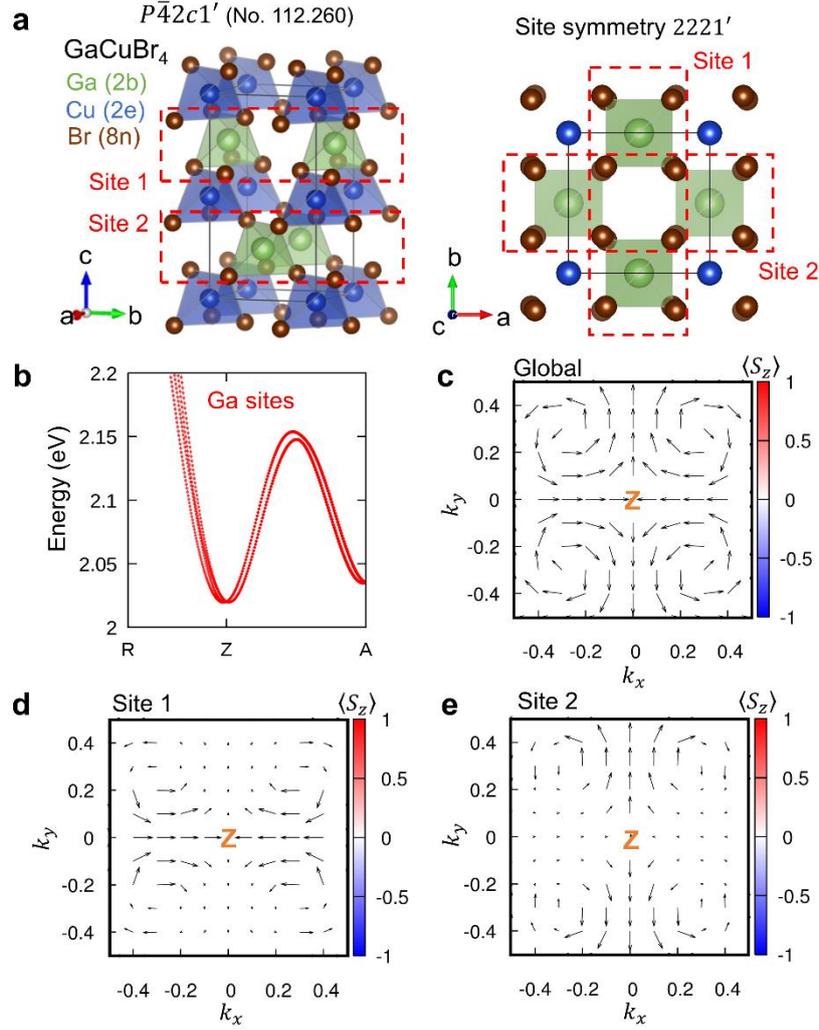

**Fig. 4 | Anisotropic spin-momentum-site locking in GaCuBr$_4$. a** Crystal structure of GaCuBr$_4$ with two GaBr$_4$ tetrahedron sites marked by the red dashed box. **b** Band structures contributed by both Ga atomic sites in GaCuBr$_4$. **c** Global spin texture in $k_z = 0.5$ plane for the band of Ga atoms. **d-e** Spin texture projected on GaBr$_4$ tetrahedron site 1 and 2, respectively.